\documentclass[prd,nofootinbib,twocolumn,showpacs]{revtex4}
\usepackage{graphics}
\usepackage{bm}

\begin{document}

\title{
Domain wall lattices
}

\author{Levon Pogosian}
\affiliation{
Theoretical Physics, The Blackett Laboratory, Imperial College,\\
Prince Consort Road, London SW7 2BZ, U.K.. }

\author{Tanmay Vachaspati}
\affiliation{Department of Physics, Case Western Reserve University,
10900 Euclid Avenue, Cleveland, OH 44106-7079, USA.}

\begin{abstract}
We construct lattices with alternating kinks and 
anti-kinks. The lattice is shown to be stable in certain
models. We consider the forces between kinks and antikinks
and find that the lattice dynamics is that of a Toda 
lattice. Such lattices are exotic metastable states in which 
the system can get trapped during a phase transition. 
\end{abstract}


\

\maketitle

Domain walls are among the simplest topological defects
known and have often been used as a test-bed for studying
non-perturbative effects. An example of a domain wall is 
the ``$\phi^4$ kink'' trivially extended to three spatial
dimensions. This solution is often thought to typify all
domain walls. However, it has recently become clear that the 
properties of $\phi^4$ kinks do not simply carry over to more 
complicated systems 
\cite{PogVac00,Vac01,PogVac01,Pog02,Davetal02,Ton02}
including condensed matter systems such as He-3 \cite{SalVol88}.
Instead a much richer structure of kinks emerges. As we show
in this paper, the enhanced structure of kinks in $SU(N)\times Z_2$ 
(for odd $N$) allows for the construction of lattices of 
kinks and anti-kinks. In related models the kink lattice 
is also perturbatively stable.

The lattices we are constructing are different from other
known lattices such as an Abrikosov lattice \cite{Abr57}.
The total topological charge of an Abrikosov lattice is 
non-vanishing. In contrast, the kink lattice we will construct
will have zero topological charge. This means that we
can construct the kink lattice in a box with periodic boundary
conditions, and also that the lattice is topologically equivalent 
to the vacuum. Hence if we start out in an unbroken symmetry phase, 
with vanishing net topological charge, there is a chance that, 
after the symmetry is spontaneously broken, the system will be 
trapped in the lattice phase instead of the true vacuum. From the 
lattice phase, the system can then only reach the true vacuum by 
quantum tunneling.

We start with an $SU(N)\times Z_2$ field theory whose
Lagrangian is:
\begin{equation}
L = {\rm Tr} (\partial_\mu \Phi )^2 - V(\Phi ) 
\label{lagrangian}
\end{equation}
where $\Phi$ is an $SU(N)$ adjoint and $V(\Phi )$ is invariant 
under
\begin{equation}
G\equiv SU(N) \times Z_2 \ . 
\label{originalsymm}
\end{equation}
$N \ge 3$ is taken to be odd, and the parameters in 
$V$ are such that $\Phi$ has an expectation value that can 
be chosen to be
\begin{equation}
\Phi_0 = \eta \sqrt{2 \over {N(N^2-1)}}
                  \pmatrix{n{\bf 1}_{n+1}&{\bf 0}\cr
                      {\bf 0}&-(n+1){\bf 1}_n\cr} \ ,
\label{phi0}
\end{equation}
where, $n = (N-1)/2$, ${\bf 1}_p$ is the $p\times p$ identity 
matrix and $\eta$ is an energy scale determined by the minima 
of the potential $V$. Such an expectation value spontaneously 
breaks the symmetry down to:
\begin{equation}
H = [SU(n+1)\times SU(n)\times U(1)]/Z_{n+1}\times Z_n 
\label{unbrokensymm} \ ,
\end{equation}
We will choose $V(\Phi )$ to be a quartic polynomial: 
\begin{equation}
V(\Phi ) = - m^2 {\rm Tr}[ \Phi ^2 ]+ h ( {\rm Tr}[\Phi ^2  ])^2 +
      \lambda {\rm Tr}[\Phi ^4]  + V_0 
\label{quarticV}
\end{equation}
where $V_0$ is a constant chosen so that the minimum of the
potential has $V=0$. The Lagrangian is symmetric under
$\Phi \rightarrow -\Phi$ and it is the breaking of this
$Z_2$ symmetry that gives rise to topological domain wall 
solutions. We could also extend the model by making it locally
gauge invariant. The solutions described below will still be
valid with the gauge fields set to zero; the stability 
analysis will change.

While our analysis can easily be carried out for general $N$, 
the physics is more transparent for a specific choice of $N$.
Hence we will choose $N=5$ and, where relevant,
remark on the case of general $N$ \footnote{The corresponding
quartic model with $N=3$ has an accidental $SO(8)$ symmetry. 
We could consider $N=3$ if (Tr$(\Phi^3))^2$ and  Tr$(\Phi^6)$
terms were added to the potential. We have chosen to work with 
quartic potentials and with the larger value of $N$.}. Then the
desired symmetry breaking to
\begin{equation}
H= [SU(3)\times SU(2)\times U(1)]/[Z_3\times Z_2] \label{His321}
\label{unbrokenH}
\end{equation}
is achieved in the parameter range
\begin{equation}
{h \over \lambda} > - {{N^2+3}\over {N(N^2-1)}}\biggl |_{N=5}
                     = - {7\over {30}} \ .
\label{symmbreakparam}
\end{equation}
The vacuum expectation value (VEV), $\Phi_0$ is (up to any
gauge rotation)
\begin{equation}
\Phi_0 = {\eta \over \sqrt{60}} {\rm diag}(2,2,2,-3,-3) 
\label{su5phi-}
\end{equation}
with $\eta \equiv {{m} / {\sqrt{\lambda '}}}$ and
\begin{equation}
\lambda ' \equiv
      h +  {{N^2+3}\over {N(N^2-1)}}\biggl |_{N=5} \lambda
      = h + {7\over {30}} \lambda \ .
\label{lambdaprime}
\end{equation}

In Refs. \cite{PogVac00,Vac01,PogVac01} it was found that there are
several domain wall solutions in this model but a solution
with least energy is achieved if 
$\Phi (-\infty) \equiv \Phi_- = \Phi_0$ and 
\begin{equation}
\Phi (+\infty) \equiv \Phi_+ =
- {\eta \over \sqrt{60}}{\rm diag}(2,-3,-3,2,2) 
\label{su5phi+}
\end{equation}
Two features of $\Phi_+$ are worthy of note. First,
there is a minus sign in front. This puts $\Phi_+$ 
and $\Phi_-$ in disconnected parts of the vacuum
manifold. The second feature is that two blocks of entries
of $\Phi_+$ are permuted with respect to those of
$\Phi_-$. In other words, $\Phi_-$ and $-\Phi_+$
are related by a non-trivial gauge rotation.  Furthermore,
the kink solution (or, domain wall solution, in more
than one dimension) can be written down 
explicitly in the case when $h/\lambda =-3/20$
\cite{PogVac00,Vac01}: 
\begin{equation}
\Phi_k = {{1-\tanh(\sigma x)}\over 2} \Phi_- +
                {{1+\tanh(\sigma x)}\over 2} \Phi_+
\label{qeq2solution}
\end{equation}
where $\sigma = m/\sqrt{2}$.
{}For other values of the coupling constants, the
solution has been found numerically \cite{PogVac00}.

The topological charge of a kink can be defined as
\begin{equation}
Q = {{\sqrt{60}}\over \eta}(\Phi_R - \Phi_L)
\end{equation}
where $\Phi_R$ and $\Phi_L$ are the asymptotic
values of the Higgs field to the right ($R$) and
left ($L$) of the kink. (The rescaling has been
done for convenience.) Then the charge of the kink
in eq. (\ref{qeq2solution}) is:
\begin{equation}
Q^{(1)} = {\rm diag}(-4,1,1,1,1)
\label{Q1}
\end{equation}
Similarly, one can construct kinks with charge
matrices $Q^{(i)}$ ($i=1,...,5$) which have $-4$ as 
the $ii$ entry and $+1$ in the remaining diagonal 
entries. Hence there are kink solutions with 5
different topological charge matrices. Individually,
the kinks can be gauge rotated into one another. But
when two kinks are present, the different charges 
are physically relevant. This is most easily seen
by noting that the interaction between a kink with
charge $Q^{(i)}$ and an antikink with charge 
${\bar Q}^{(j)}= - Q^{(j)}$ is proportional to 
${\rm Tr}(Q^{(i)}{\bar Q}^{(j)})$ \cite{Pog02}.
Then we have
\begin{eqnarray}
{\rm Tr}(Q^{(i)}{\bar Q}^{(j)}) 
               &=& -20 \ {\rm if} \ i=j \nonumber \\
               &=& +5  \ {\rm if} \ i \ne j
\label{Qtraces}
\end{eqnarray}
The sign of the trace tells us if the force between the kink
and antikink is attractive (minus) or repulsive (plus).
Hence the force between a kink and an antikink with different
orientations ($i\ne j$) is repulsive. This observation is key 
to the construction of kink lattices.

In Ref. \cite{Pog02}, the repulsive potential between a kink 
and an antikink at rest was derived. When the kink and
antikink separation, $R$, is large, the result reduces to:
\begin{equation}
U(R) = {{4\sqrt{2}m^3}\over {\lambda}} e^{-2\sqrt{2}mR}
\label{U(R)}
\end{equation}

To construct a kink lattice, we now need to arrange
a periodic sequence of kink charges such that the nearest
neighbor interactions are repulsive. Kinks that are not
nearest neighbors but are further apart will also interact,
and perhaps even attract each other. However the forces
between kinks and antikinks fall off exponentially fast
and just taking nearest-neighbor interactions into account
should be sufficient, at least for lattice spacing larger
than the kink width. So now we can write down a sequence of
charges that can form a kink lattice. This is:
\begin{equation}
...
Q^{(1)}{\bar Q}^{(5)}Q^{(3)}{\bar Q}^{(1)}Q^{(5)}{\bar Q}^{(3)}
...
\label{minlatt}
\end{equation}
and the sequence just repeats itself. Alternately, we could have
a finite lattice if the kinks were in a compact space, such as
a compact higher dimension, or the $S^1$ that arises in evaluating
the partition function in statistical mechanics. 

The sequence listed above is the minimum sequence for which
the nearest neighbor interactions are repulsive. The repeating
length of 6 kinks is independent of $N$ in $SU(N)$ since it is clear
that we need at least, and no more than, 3 different kinds of
kink charges.

Another way to write the kink sequence is to write it
as a sequence of Higgs field expectation values. We write
this sequence for the above minimal lattice:
\begin{eqnarray}
... &\rightarrow& +(2,2,2,-3,-3)
              \rightarrow -(2,-3,-3,2,2) \nonumber \\
              &\rightarrow& +(-3,2,2,-3,2)
              \rightarrow -(2,-3,2,2,-3) \nonumber \\
              &\rightarrow& +(2,2,-3,-3,2)
              \rightarrow -(-3,-3,2,2,2) \nonumber \\
              &\rightarrow& +(2,2,2,-3,-3)
                                 \rightarrow ...
\label{Higgs sequence}
\end{eqnarray}

\begin{figure}
\scalebox{0.40}{\includegraphics{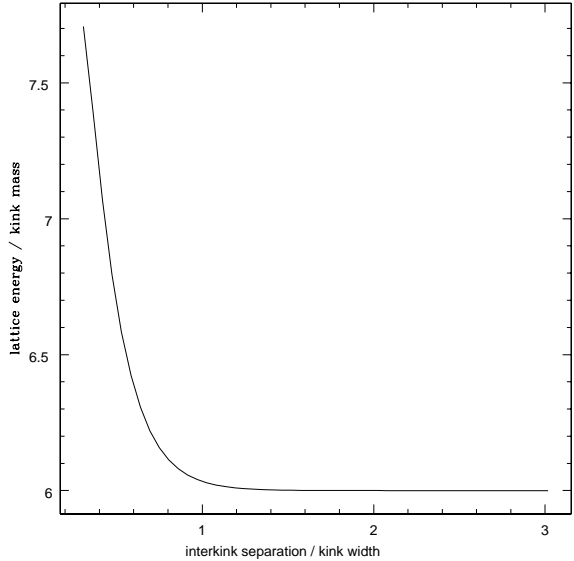}}
\caption{\label{energyvsspacing}
The energy of the minimal lattice versus
lattice spacing for $h/\lambda = -3/20$, $\lambda=0.5$ 
and $\eta=1$.
}
\end{figure}

We have constructed the solution for the minimal kink lattice
numerically on a space with periodic boundary conditions.
In Fig. \ref{energyvsspacing} we show the total energy of the
minimal lattice as a function of lattice spacing.

The minimal lattice of 6 kinks is easily generalized
to longer sequences. A sequence of 10 kinks in the $N=5$ case
is aesthetic in the sense that it uses all the 5 different
charge matrices democratically:
\begin{equation}
...
Q^{(1)}{\bar Q}^{(5)}Q^{(3)}{\bar Q}^{(4)}Q^{(2)}
{\bar Q}^{(1)}Q^{(5)}{\bar Q}^{(3)}Q^{(4)}{\bar Q}^{(2)}
...
\label{10latt}
\end{equation}
Similarly one can construct sequences in the general $N$
case.

We have also numerically studied the dynamics of the
lattice by giving one of the kinks an initial velocity.
We find that the kink scatters elastically on the
neighboring anti-kink, and the motion propagates down
the lattice. Indeed, lattices of masses interacting
via exponentially decaying repulsive forces (see eq. (\ref{U(R)}))
have been studied in the literature and are known as 
Toda lattices \cite{Tod88}. Hence the kink lattice is
a Toda lattice.

\begin{figure}
\scalebox{0.40}{\includegraphics{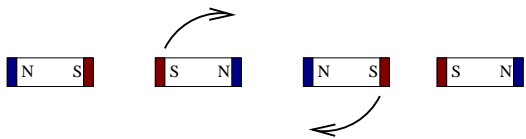}}
\caption{\label{magnets}
A linear row of bar magnets placed North to North and
South to South has an instability towards rotations in
the transverse directions as shown. 
}
\end{figure}

We now discuss the stability of the lattice. A detailed
stability analysis shows that the lattice in eq. (\ref{minlatt})
has three unstable modes, corresponding to rotations in the 1-3,
1-5, 3-5 blocks. 
To clarify the instability, we draw an analogy between the 
kink lattice and a lattice of bar magnets (Fig. \ref{magnets}).
Each bar magnet on its own has a zero mode that corresponds
to rotations in three dimensional space. When placed in
the lattice shown in Fig. \ref{magnets}, rotational
zero modes turn into unstable modes. Similarly, an
isolated kink has zero modes corresponding to rotations
in gauge space -- for example, a kink with charge $Q_1$
can be rotated into the kink with charge $Q_3$ without
any cost in energy. When a kink of charge $Q_1$ is placed 
near an antikink of charge ${\bar Q}_3$, the zero mode
becomes an unstable mode, making it favorable for $Q_1$
to rotate into $Q_3$ after which the kink and antikink
can annihilate.

This understanding of the instability also suggests a 
resolution: if the rotational zero modes are sufficiently
suppressed or absent, the lattice will become stable. To 
suppress the rotational zero modes, we could break the 
symmetry further so that the kink is no longer invariant 
under rotations. We can also consider a case where the 
zero modes are completely absent right from the start.
We will discuss this latter case as it is simpler to deal 
with and provides an explicit example of a field theory 
with stable kink lattices. 

Consider the model with four real scalar fields,
\begin{equation}
L = {1\over 2}\sum_{i=1}^4 (\partial_\mu f_i) ^2
     + V(f_1,f_2,f_3,f_4)
\end{equation}
and
\begin{eqnarray}
V &=&
-{m^2\over 2}\sum_{i=1}^4 f_i^2
+ {h \over 4}(\sum_{i=1}^4 f_i^2)^2
+ {\lambda \over 8} \sum_{a=1}^3 f_a^4
\nonumber \\
&+& {\lambda \over 4} \left[ 
{7\over 30} f_4^4 + f_1^2 f_2^2 \right]
+ {\lambda \over {20}} [4(f_1^2 + f_2^2) + 9f_3^2] f_4^2
\nonumber \\
&+& {\lambda \over \sqrt{5}}f_2 f_4 \left( f_1^2
- {f_2^2 \over 3} \right) + {m^2 \over 4} \eta^2
\end{eqnarray}
This model has been obtained by truncating the field $\Phi$
occurring in eq. (\ref{lagrangian}) to its diagonal elements.
The fields $f_1$ and $f_2$ correspond to the diagonal
generators $\lambda_3$ and $\lambda_8$ of $SU(3)$ 
(see eq. (\ref{unbrokenH})) in
the Gell-Man basis, $f_3$ corresponds to the diagonal
generator $\tau_3$ of $SU(2)$, and $f_4$ corresponds 
to the generator of $U(1)$.
Now our four field model does not have the continuous 
$SU(5)$ symmetry of the model in eq. (\ref{lagrangian}).
The only remnant of the $SU(5)$ symmetry corresponds 
to the permutation of the five diagonal entries of $\Phi$. 
In addition, the model also has the $Z_2$ symmetry under 
which $f_i \rightarrow -f_i$. Hence the model has an
$S_5\times Z_2$ symmetry. 

A vacuum of the model is given by $f_1=0=f_2=f_3$ and
$f_4 \ne 0$. This breaks the symmetry to 
$S_3\times S_2$, corresponding to permutations of $\Phi$
in the $SU(3)$ and $SU(2)$ blocks. The vacuum manifold
consists of $5!\times 2/3!\times 2! = 20$ discrete points.
If we fix the vacua at $x=-\infty$, this implies that
there are 20 kink solutions in the model. All these
20 kink solutions have been described in Ref. \cite{PogVac01}.

The construction of kink lattices proceeds exactly as
in the $SU(5)$ case above because the off-diagonal
components of $\Phi$ vanish there. Hence the $S_5\times Z_2$
model contains kink lattice solutions. Furthermore,
these lattices are stable because the dangerous
rotational perturbations are absent by the very
construction of the model.

The occurrence of stable kink lattices with net vanishing
topological charge implies that there are metastable states
in the field theory. Generally metastable states are
present in field theories due to features in the potential.
Here, however, the metastable states are non-perturbative
features of the model.

\begin{figure}
\vskip 0.2 truecm
\scalebox{0.40}{\includegraphics{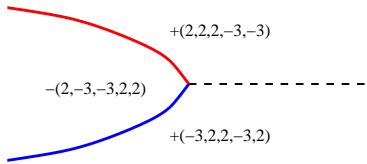}}
\caption{\label{wall-junction}
The distribution of Higgs expectation values in three
domains can lead to a wall junction \cite{Davetal02}. 
In the $SU(5)$ model, the dashed line is a non-topological 
wall \cite{PogVac01}. In the $S_5$ model the dashed 
line denotes a topological wall but without $Z_2$ charge.
}
\end{figure}

The existence of domain wall lattices is of interest
in the context of phase transitions. What is the
probability that a domain wall lattice will form 
during a phase transition? The answer depends on the 
complicated dynamics of a domain wall network in three 
spatial dimensions. For example, the model admits wall junctions
of the kind shown in Fig. \ref{wall-junction} \cite{Davetal02}
and different walls can have different tensions. A
simpler situation to consider is the formation 
in one spatial dimension for the $S_5$ model. We first
note that the kinks with charge given in eq. (\ref{Q1})
(and permutations thereof) are the lightest kinks in the
system having $Z_2$ topology. Other kinks will decay 
into these kinks upon evolution. So we can restrict
our attention to a sequence of kinks with charges
given in eq. (\ref{Q1}) and permutations. Now 
let us assume that we have a kink with charge $Q_1$. 
A neigboring antikink can have charge ${\bar Q}_1$, 
${\bar Q}_4$ or ${\bar Q}_5$ (see eq. (\ref{Higgs sequence})).
Of these only the first is unsuitable for a lattice
and has a probability 1/3.
Therefore if the phase transition produces $2n$ kinks,
then the probability of having exactly $2j$ kinks that
annihilate and $2n-2j$ survive to form a lattice is 
derived by finding the number of ways of choosing the
$j$ annihilating pairs and $n-j$ surviving pairs and
multiplying by the probability of annihilation (1/3)
and survival (2/3). The result is that the probability
of exactly $j$ pairs annihilating is:
$^n C_j (1/3)^j (2/3)^{n-j}$.
Summing this expression from $j=0$ to $n-3$ gives the 
total probability for obtaining a lattice provided we
have $2n$ kinks. The sum can easily be evaluated. 
The interesting limit is when $2n$ is large. In that case, 
the probability tends to unity. Hence a kink lattice is 
certain to form if there are a large number of kinks.
Further, the number of kinks is large if a large number 
of correlation domains are produced during the phase 
transition.

It would be interesting to test these ideas in a laboratory
systems in which a kink lattice can exist. Periodic boundary
conditions could be achieved if a toroidal sample were to
undergo a phase transition.

Finally we mention the implications of a domain wall lattice
produced during a cosmological phase transition. If
spacetime is $R^4\times S^1$ and the wall lattice resides
in the (small) compact dimension, there will be an effective
cosmological constant in the $R^4$ due to invariance
under Lorentz boosts of the wall Ref. \cite{Zeletal74,Vil81}. 
The effective cosmological constant may be time
dependent if the coupling constant $\lambda$ were to
run with energy scale, or to depend on the dynamics of 
the spacetime, or on another field. Yet another source of 
time dependence can come via the number of walls in the
lattice since the wall lattice is not protected
by topology or any conserved number. So the number of 
walls in the lattice can cascade down and eventually 
become zero. The difficulty with this cosmological scenario
is that the extra compact dimension will not be
static and will lead to an effective Newton's 
gravitational constant that is time dependent. Since
the metric of the system is not yet known, it is not
possible to say if the time variation can be slow
enough for the scenario to be viable.  

In conclusion, we have shown that stable lattices of
domain walls can exist in a wide class of field theories.
These are exotic metastable states in which the system
can get trapped with high probability during a phase
transition. 

\begin{acknowledgments} 
TV is grateful to Craig Copi, Jaume Garriga, Tom Kibble, 
Arthur Lue, Harsh Mathur, Glenn Starkman, and Alex Vilenkin 
for discussions. We thank the organizers of the 2002 ESF 
COSLAB School in Cracow where a part of this project was 
completed. TV was supported by DOE grant number DEFG0295ER40898 
at CWRU.

\end{acknowledgments}

\end{document}